\documentclass[floats, prd, eqnum, showpacs, nofootinbib, 
preprint,
eqsecnum]{revtex4-1}

\usepackage{color,graphicx}
\usepackage{amsfonts}
\usepackage{amssymb}
\begin{document}

\title{High energy particle collisions in static, spherically symmetric black-hole-like wormholes}
\author{Naoki Tsukamoto${}^{1}$}\email{tsukamoto@rikkyo.ac.jp}
\author{Takafumi Kokubu ${}^{2}$}\email{kokubu@hunnu.edu.cn, 14ra002a@al.rikkyo.ac.jp}
\affiliation{
${}^{1}$Department of General Science and Education, National Institute of Technology, Hachinohe College, Aomori 039-1192, Japan \\
${}^{2}$Department of Physics and Synergetic Innovation Center for Quantum Effects and Applications, Hunan Normal University, Changsha, Hunan 410081, China \\
}

\begin{abstract}
A Damour-Solodukhin wormhole with a metric which is similar to a Schwarzschild black hole
seems to be a black hole mimicker 
since it is difficult to distinguish them by practical astrophysical observations. 
In this paper, 
we investigate a center-of-mass energy for the collision of two test particles in the Damour-Solodukhin
wormhole spacetime. 
We show that the center-of-mass energy for the head-on collision of the particles is large 
if the difference between the 
metrics of the wormhole and the black hole is small.
To deeply understand the high energy particle collision, 
we generalize the head-on collision to static, spherically symmetric black-hole-like wormholes.
\end{abstract}

\maketitle

\section{Introduction}
Black holes predicted in general relativity are accepted as common compact objects in astrophysics 
after the detections of gravitational waves from binary black hole systems reported by LIGO and VIRGO collaborations~\cite{Abbott:2016blz,LIGOScientific:2018mvr}.
The detection of the shadow of a supermassive black hole candidate in the center of a giant elliptical galaxy M87 
has been reported by Event Horizon Telescope Collaboration~\cite{Akiyama:2019cqa}.
The observed shadow is explained by the presence of the Kerr black hole solution, which is a vacuum solution of Einstein equations,
but it does not exclude the possibility of alternative compact objects with a metric similar to a black hole metric coincidentally as discussed in Refs.~\cite{Akiyama:2019cqa,Damour:2007ap,Broderick:2015tda}.  

General relativity does
not prohibit spacetimes with nontrivial topological structures such as traversable wormhole spacetimes~\cite{Visser_1995,Morris_Thorne_1988}, which
are known as the solution of the Einstein equations.
One might consider that the wormholes must be prohibited since
the weak energy condition must be violated at the throat of any static and spherically symmetric traversable wormhole in general relativity~\cite{Morris_Thorne_1988}.  
However, we can make 
wormhole solutions without the violation of the weak energy condition in alternative gravitational theories~\cite{Shaikh:2016dpl}. 

The hunting of wormholes in nature can be one of interesting topics in general relativity and alternative gravitational theories to understand our Universe.
The upper bound of the number density of wormholes in the Universe is investigated in Ref.~\cite{Takahashi_Asada_2013}
with gravitational lensing~\cite{Schneider_Ehlers_Falco_1992,Schneider_Kochanek_Wambsganss_2006,Perlick_2004_Living_Rev}.
Observations of phenomena in strong gravitational fields with a high accuracy will help us to distinguish wormholes from black holes 
since the wormholes and black holes are characterized by a throat and an event horizon, respectively.
Thus, the phenomena related with light rays passing through the wormhole throat in the strong gravitational field,
such as visualizations~\cite{Muller_2004,James:2015ima}, 
shadows~\cite{Ohgami:2015nra,Ohgami:2016iqm,Kuniyasu:2018cgv,Paul:2019trt}, 
deflection angles~\cite{Chetouani_Clement_1984},
gravitational lensing~\cite{Perlick_2004_Phys_Rev_D,Nandi:2006ds,Tsukamoto:2016zdu,Shaikh:2018oul,Shaikh:2019jfr}, and 
wave optics~\cite{Nambu:2019sqn},
have been investigated eagerly.

Damour and Solodukhin considered a wormhole with 
a metric which is similar to 
the Schwarzschild black hole~\cite{Damour:2007ap}.
They concluded that we cannot distinguish the black holes from the wormholes with practical and astrophysical observations on a limited timescale $\Delta t$, 
which is a coordinate time, 
if their metrics are similar each other.
Lemos and Zaslavskii calculated the Riemann tensor in an orthonormal frame and a tidal force acting on a falling body in the Damour-Solodukhin wormhole spacetime~\cite{Lemos:2008cv}.
They pointed out that the tidal force near the wormhole is larger than the one near the Schwarzschild black hole 
and that the wormhole and the black hole can be distinguished if the falling body emits light rays near them.
Recently, emissions from its accretion disk~\cite{Karimov:2019qco},
quasinormal modes~\cite{Volkel:2018hwb,Ovgun:2019yor},
images of its accretion disks~\cite{Paul:2019trt}, 
the shadow image~\cite{Amir:2018pcu},
and gravitational lensing~\cite{Nandi:2018mzm,Ovgun:2018fnk,Ovgun:2018oxk} 
in the Damour-Solodukhin wormhole spacetime have been investigated.
Gravitational-wave echoes by the wormhole~\cite{Bueno:2017hyj} have also investigated. 
See Refs.~\cite{Cardoso:2019apo,Li:2019kwa,Bronnikov:2019sbx} and the references therein for the echoes.

Piran~\textit{et al.} found that the center-of-mass energy of the rear-end collision of two test particles can be arbitrarily high near an extremal Kerr black hole~\cite{Piran_1975,Piran:1977dm}.
The particle collision is often called the Ba\~{n}ados-Silk-West~(BSW) mechanism or BSW collision
since Ba\~{n}ados~\textit{et al.} rediscovered that in 2009~\cite{Banados:2009pr}.
For the arbitrarily high center-of-mass energy, the conserved angular momentum of either of the particles must be fine tuned, 
and the 
particle falls into the 
black hole with an infinite proper time.
Since the rediscovery,
the details of the rear-end particle collision have been investigated eagerly~\cite{Harada:2014vka,Zakria:2015eua}.

The center-of-mass energy of the head-on collision of two particles can be arbitrarily large in a near-horizon limit in the Kerr and Schwarzschild spacetimes~\cite{Piran:1977dm,Zaslavskii:2019bho}.~\footnote{In 
Ref.~\cite{Baushev:2008yz}, the upper bound of the center-of-mass energy of two falling particles 
which are at rest at infinity with an equal mass $m$ is obtained as $E_{cm}=2\sqrt{5}m$ in the Schwarzschild spacetime.}
However, particles which fall into a black hole do not cause the head-on collision directly. 
We need to seek an astrophysical scenario that outgoing particles can exist near the event horizon to cause the head-on collision.

On the other hand, particles can collide with each other head on directly
in spacetimes without the event horizon.
Patil~\textit{et al.} have shown an arbitrarily high center-of-mass energy of a head-on particle collision within a finite proper time of two particles
in the overspinning and near-extremal Kerr spacetime~\cite{Patil:2011ya}
and in an overcharged and near-extremal Reissner-Nordstr\"{o}m spacetime~\cite{Patil:2011uf} with naked singularity.
Zaslavskii has generalized the head-on collision in the near-extremal spacetimes without the event horizon~\cite{Zaslavskii:2013nra},
but the method works in near-extremal spacetimes only.
Krasnikov has constructed a wormhole metric and 
found a center-of-mass energy for the head-on collision of particles at a throat can be arbitrarily high in a black hole limit~\cite{Krasnikov:2018nga}.
A head-on particle collision near a white hole was investigated in Ref.~\cite{Zaslavskii:2017guu}.

In this paper, we investigate the collisions of two test particles in the Damour-Solodukhin wormhole spacetime.
We show that the center-of-mass energy of a head-on collision at a throat is large
if the metric of the wormhole spacetime is similar to the Schwarzschild spacetime.
 
We generalize the particle collision to a general, static, and spherically symmetric wormhole case.
We show that the center-of-mass energy of the head-on collision can be large 
when the metrics of the wormholes are similar to black holes.
As additional examples, we consider a Reissner-Nordstr\"{o}m black-hole-like wormhole~\cite{Lemos:2008cv}, 
Krasnikov's wormhole~\cite{Krasnikov:2018nga}, and the Ellis wormhole~\cite{Ellis_1973,Bronnikov_1973}.

This paper is organized as follows. 
We investigate the particle collision in the Damour-Solodukhin wormhole spacetime in Sec.~II 
and in a general, static, and spherically symmetric wormhole spacetime in Sec.~III.
In Sec.~IV, we discuss and conclude our results.
In Appendix~A, we consider a relation of the center-of-mass energy of the particle collision and the Ricci scalar in the wormhole spacetimes.
In this paper, we use the units in which a light speed and Newton's constant are unity.

\section{Damour-Solodukhin wormhole spacetime}
In this section, we review a particle motion in the Damour-Solodukhin spacetime and we investigate the collision of two particles there. 
We also consider the collision of a particle with a circular motion at a throat and another particle.

\subsection{Particle motion}
The metric of the Damour-Solodukhin wormhole spacetime~\cite{Damour:2007ap} with the throat at $r=r_0 \equiv 2M$ is given by 
\begin{eqnarray}\label{eq:line_element1} 
ds^2
&=&-\left( f(r)+\Lambda^2 \right)dt^2 +\frac{dr^2}{f(r)} \nonumber\\
&&+r^2 \left( d \theta^2 +\sin^2\theta d \phi^2 \right),
\end{eqnarray}
where $f(r)$ is defined by $f(r)\equiv 1-2M/r$, 
$M$ is a positive mass parameter, 
$\Lambda$ is a small and positive parameter, 
and a radial coordinate $r$ is defined in a range $2M \leq r < \infty$.
There are the time-translational and axial Killing vectors 
$t^\mu \partial_\mu=  \partial_t$ and $\phi^\mu \partial_\mu = \partial_\phi$ 
because of the stationarity and axisymmetry of the spacetime, respectively. 
We concentrate on a particle motion on an equatorial plane $\theta=\pi/2$ in this paper.

A particle has a conserved energy $E\equiv -g_{\mu \nu}t^\mu p^\nu=-g_{tt}p^t=(f(r)+\Lambda^2)p^t$, where $p^\mu$ is the 4-momentum of the particle, 
and a conserved angular momentum $L\equiv g_{\mu \nu} \phi^\mu p^\nu=g_{\phi\phi}p^\phi=r^2 p^\phi$.
From the conserved energy $E$, the conserved angular momentum $L$ and the 4-momentum $p^\mu=dx^\mu/d\lambda$, where $\lambda$ is a parameter along the world line of the particle, 
we obtain
\begin{eqnarray}
\frac{dt}{d\lambda}    &=&  p^t   = \frac{E}{f(r)+\Lambda^2}, 
\label{eq:dt} \\
\frac{d\phi}{d\lambda} &=& p^\phi = \frac{L}{r^2}.
\end{eqnarray}
From $p^\mu p_\mu =-m^2$, where $m$ is the mass of the particle, the equation of the radial motion of the particle is given by 
\begin{eqnarray}\label{eq:r_motion1}
\left(
\frac{dr}{d \lambda} 
\right)^2
+V(r)=0, 
\end{eqnarray}
where $V(r)$ is an effective potential for the radial motion defined by 
\begin{eqnarray}\label{eq:V} 
V(r) \equiv f(r) \left( m^2- \frac{E^2}{f(r)+\Lambda^2} +\frac{L^2}{r^2} \right). 
\end{eqnarray}
The particle can be in a region where the effective potential $V(r)$ is nonpositive.
Note that the particle with $E\geq m\sqrt{1+\Lambda^2}$ can exist at a spacial infinity.
From Eq.~(\ref{eq:r_motion1}), we obtain 
\begin{eqnarray}\label{eq:pr} 
\frac{dr}{d\lambda}
=p^r
=\sigma \sqrt{-V(r)}, 
\end{eqnarray}
where $\sigma$ is defined by $\sigma =1$ $(-1)$ for an outgoing (ingoing) particle.

The radial coordinate $r$ is not defined well at the throat $r=2M$~\cite{Damour:2007ap}.
To cover the wormhole spacetime globally
we introduce a proper radial coordinate $\rho$ 
defined by 
\begin{eqnarray}\label{eq:rho}
\left| \rho \right|
&=&\int^{r}_{2M}\frac{dr}{\sqrt{f(r)}} \nonumber\\
&=&r\sqrt{f(r)}+M \log \left[ -1 +\frac{r}{M} +\frac{r}{M} \sqrt{f(r)} \right]
\end{eqnarray}
in the range $-\infty< \rho < \infty$~\footnote{The proper radial coordinate $\rho$ is denoted by $y$ in Ref.~\cite{Damour:2007ap}.}.
Using the proper radial coordinate $\rho$,
the line element~(\ref{eq:line_element1}) is rewritten as
\begin{eqnarray}\label{eq:line_element2} 
ds^2
=-\left( f(\rho)+\Lambda^2 \right)dt^2 +d\rho^2 +r^2(\rho) \left( d \theta^2 +\sin^2\theta d \phi^2 \right), \nonumber\\
\end{eqnarray}
where $f(\rho)$ is given by
\begin{eqnarray}
f(\rho)=f\left(r(\rho)\right)=1-\frac{2M}{r(\rho)}.
\end{eqnarray}
The wormhole throat is located at $\rho=0$.
The equation of the radial motion is rewritten as
\begin{eqnarray}\label{eq:rho_motion}
\left(
\frac{d\rho}{d \lambda} 
\right)^2
+v(\rho)=0, 
\end{eqnarray}
where $v(\rho)$ is an effective potential for the radial motion in the radial coordinate $\rho$ given by 
\begin{eqnarray}\label{eq:v}
v(\rho) \equiv m^2- \frac{E^2}{f(\rho)+\Lambda^2} +\frac{L^2}{r^2(\rho)}
\end{eqnarray}
and the particle can exist in a region where $v(\rho)$ is nonpositive.
When $\Lambda$ is small, the bump of the effective potential at the throat, which is given by $v(0)=m^2-E^2/\Lambda^2+L^2/(4M^2)$, is deep,
while the effective potential in the radial coordinate $r$ at the throat is given by $V(2M)=0$.
Figure 1 shows the dimensionless effective potential $v/m^2$ of a particle with a specific conserved energy~$e \equiv E/m = \sqrt{1+\Lambda^2}$
and vanishing conserved angular momentum $L=0$ as the function of a dimensionless radial proper distance $\rho/M$.
At the throat, the dimensionless effective potential of the particle is obtained as $v(0)/m^2=-1/\Lambda^2$.
\begin{figure}[htbp]
\begin{center}
\includegraphics[width=87mm]{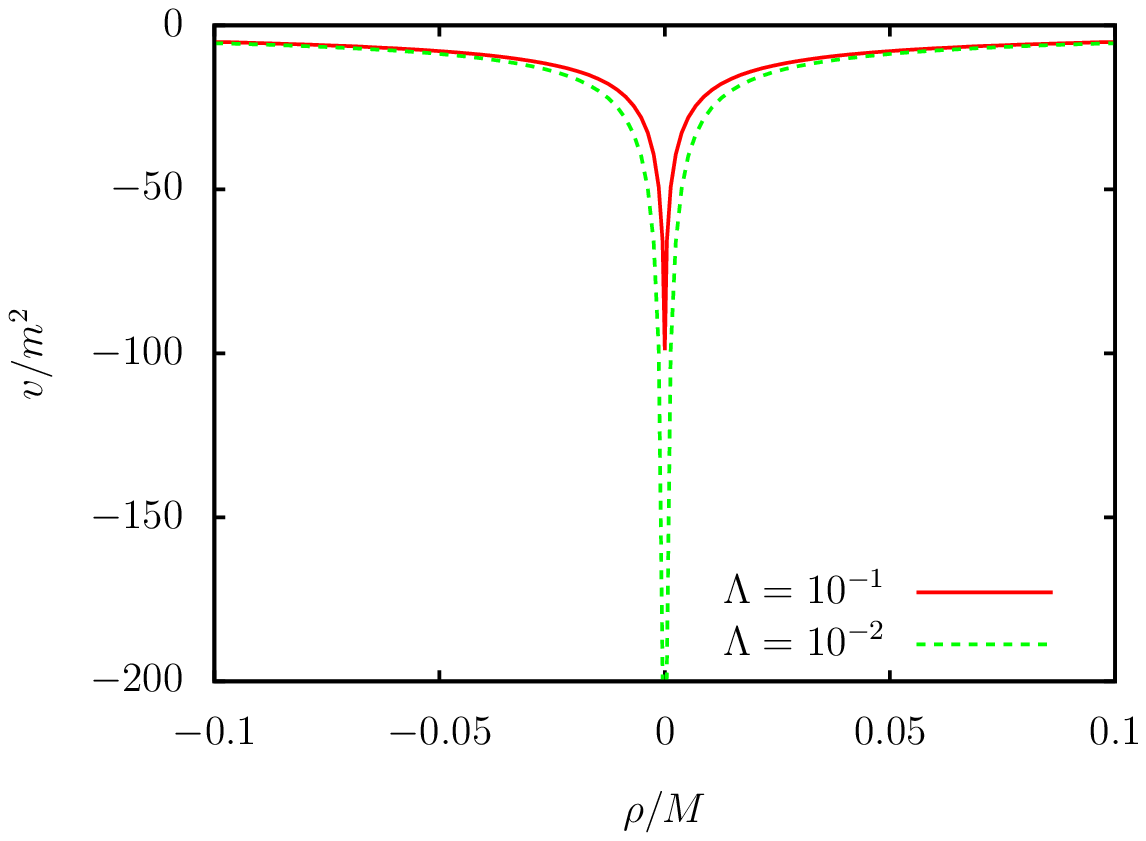}
\end{center}
\caption{The examples of the dimensionless effective potential $v/m^2$ as the function of a dimensionless radial proper distance $\rho/M$ 
for the radial motion of the particle 
with $e\equiv E/m= \sqrt{1+\Lambda^2}$
and $L=0$.
The solid~(red) and dashed~(green) curves denote
the dimensionless effective potentials $v(\rho/M)/m^2$ in the wormhole spacetime with $\Lambda=10^{-1}$ and $10^{-2}$, respectively.
There is a deep bump of the dimensionless effective potential with $v(0)/m^2=-1/\Lambda^2$ at the throat $\rho/M=0$.}
\label{fig:v}
\end{figure}

Under assumptions $e \geq \sqrt{1+\Lambda^2}$, $\Lambda^2 \ll 1$, and $L=0$, 
we consider a coordinate time $\Delta t$ 
that a particle reaches to the throat at $r=0$ from $r=l>2M$.
The coordinate time $\Delta t$ is obtained as
\begin{eqnarray}\label{eq:Delta_t}
\Delta t 
= e \int^{l}_{2M} \frac{dr}{\sqrt{f(r)\left(f(r)+\Lambda^2\right)} \sqrt{e^2-f(r)-\Lambda^2}}. \nonumber\\
\end{eqnarray}
We also assume that the particle almost stops but moves slowly near the throat from the point of view of a distant observer, in other words, it is frozen there, 
and that the coordinate time $\Delta t$ is dominated not by $l$ but by $\Lambda$.\footnote{From Eq. (\ref{eq:Delta_t2}), 
we notice that a condition $l \ll \Delta t \sim -4M\log \Lambda$ must be satisfied so that the coordinate time $\Delta t$ is dominated by $\Lambda$.}
Under the assumptions, the coordinate time $\Delta t$ is given by
\begin{eqnarray}\label{eq:Delta_t2}
\Delta t 
&\sim& \int^{l}_{2M} \frac{dr}{\sqrt{f(r)\left(f(r)+\Lambda^2\right)}} \nonumber\\
&\sim& -4M \log \Lambda.
\end{eqnarray}
In the black hole limit $\Lambda \rightarrow 0$, the coordinate time $\Delta t$ diverges.
Equation~(\ref{eq:Delta_t2}) is rewritten as
\begin{eqnarray}\label{eq:Lambda}
\Lambda \sim \exp{\left(-\frac{\Delta t}{4M}\right)}.
\end{eqnarray}

As Damour and Solodukhin assumed~\cite{Damour:2007ap},
we assume that matters near the wormhole started to fall into a wormhole $\Delta t =6\times 10^9$ years ago.
If $\Lambda \ll \exp \left( -10^{18} \right)$ and if the mass of the wormhole is $M=6.5\times 10^9 M_{\odot}$~\footnote{The mass $M=3\times 10^6 M_{\odot}$ 
has been used in Ref.~\cite{Damour:2007ap}, 
while we use $M=6.5\times 10^9 M_{\odot}$, 
which is the same mass 
as the mass of a supermassive black hole candidate in M87~\cite{Akiyama:2019cqa}.}, from Eq.~(\ref{eq:Delta_t}),
the falling matters cannot pass the throat, 
and they cannot be reflected by the throat 
because they are almost frozen near the throat.
Thus, an observer cannot distinguish the wormhole from a black hole with the same mass as the mass of the wormhole. 
On the other hand, if $\exp \left( -10^{18} \right) \lesssim  \Lambda \ll 1$ is satisfied,
we could distinguish the wormhole from the black hole with observations near the throat such as a shadow 
because of outgoing flows from the throat in principle.

The proper time $\tau \equiv m \lambda$ of the particle with $E> m\sqrt{1+\Lambda^2}$, $0<\Lambda \ll 1$, and $L=0$ initially at $r=l>2M$ to reach the throat at $r=2M$ 
is given by, from Eqs.~(\ref{eq:V}) and (\ref{eq:pr}), 
\begin{eqnarray}
\tau 
\sim \int^{l}_{2M} \sqrt{\frac{r}{2M}}dr 
\sim \frac{l\sqrt{2l}}{3\sqrt{M}}.
\end{eqnarray}
The proper time does not diverge even in the black hole limit $\Lambda \rightarrow 0$.

\subsection{Center-of-mass energy of two particles}
We consider that particles $1$ and $2$ collide at a point in the Damour-Solodukhin wormhole spacetime.
The center-of-mass energy $E_{cm}(r)$ of the particles at the collisional point is given by 
\begin{eqnarray}
E_{cm}^2(r)
&\equiv&-g_{\mu \nu} \left( p^{\mu}_{1}+p^{\mu}_{2} \right) \left( p^{\nu}_{1}+p^{\nu}_{2} \right) \nonumber\\
&=&m_1^2+m_2^2+\frac{2 E_1 E_2 }{f(r)+ \Lambda^2 } \nonumber\\
&&-\frac{2  \sigma_1 \sigma_2 \sqrt{V_1(r)V_2(r)}}{f(r)} -\frac{2 L_1 L_2}{r^2}\nonumber\\
&=&m_1^2+m_2^2+\frac{2 E_1 E_2 }{f(r)+ \Lambda^2 }-\frac{2 L_1 L_2}{r^2} \nonumber\\
&&-2\sigma_1 \sigma_2 \sqrt{ m_1^2- \frac{E_1^2}{f(r)+\Lambda^2} +\frac{L_1^2}{r^2}} \nonumber\\
&&\times \sqrt{m_2^2- \frac{E_2^2}{f(r)+\Lambda^2} +\frac{L_2^2}{r^2}},
\end{eqnarray}
where $p_i^\mu$, $m_i$, $E_i$, $\sigma_i$, $V_i(r)$, and $L_i$ are 
$p^\mu$, $m$, $E$, $\sigma$, $V(r)$, and $L$ for particle $i=1$ or $2$, respectively.
We set $\sigma_1 \sigma_2=-1$ and $1$ for a head-on collision and a rear-end collision, respectively.
Using the effective potential $v_i(r)=v_i(\rho(r))$ for particle $i=1$ or $2$ in the proper radial coordinate $\rho=\rho(r)$,
the center-of-mass energy $E_{cm}(r)$ is expressed as 
\begin{eqnarray}
E_{cm}^2(r)
&=&m_1^2+m_2^2+\frac{2 E_1 E_2 }{f(r)+ \Lambda^2 } \nonumber\\
&&-2  \sigma_1 \sigma_2 \sqrt{v_1(r)v_2(r)} -\frac{2 L_1 L_2}{r^2}.
\end{eqnarray}
The center-of-mass energy of the collision is large when the relative velocity of the colliding particles is large~\cite{Zaslavskii:2011dz}.

We assume that
$\Lambda$ is small and it satisfies $\Lambda \ll E_1/m_1$, $E_2/m_2$, 
$E_1M/\left|L_1 \right|$~\footnote{When $\Lambda \ll EM/\left|L\right|$ is satisfied, the third terms in the effective potentials $V(r)$~(\ref{eq:V}) and $v(\rho)$~(\ref{eq:v}) can be ignored.},
 and $E_2M/\left|L_2\right|$ in the rest of Sec.~II-B.

\subsubsection{Head-on particle collision with $\sigma_1 \sigma_2=-1$}
For a head-on collision, we set $\sigma_1 \sigma_2=-1$.
As we have shown in Sec.~II-A, the effective potential $v(\rho)$ in the proper radial coordinate $\rho$ has a deep depth at the throat.
The colliding particles have large velocities in a facing direction, 
and the relative velocity is large. 
Thus, the center-of-mass energy of the collision of the head-on particles becomes large. 
The center-of-mass energy of the particles at the throat $r=r_0=2M$ 
is given by
\begin{eqnarray}\label{eq:Ecm_headon_DS}
\frac{E_{cm}(r_0)}{\sqrt{E_1 E_2}}
\sim \frac{2}{\Lambda}.
\end{eqnarray}
It can be arbitrarily large in a black hole limit $\Lambda \rightarrow 0$ 
as well as the center-of-mass energy of a head-on particle collision in a near-horizon limit in the Schwarzschild spacetime~\cite{Zaslavskii:2019bho}.

We consider a wormhole with a mass $M=6.5\times 10^9 M_{\odot}$
and into which particles started to fall~$\Delta t=6\times 10^9$ years ago.
From observations, 
we would distinguish a wormhole with $\exp \left( -10^{18} \right) \lesssim  \Lambda \ll 1$ from a black hole with the same mass in principle.
The center-of-mass energy for the head-on particle collision is given by $1 \ll E_{cm}/\sqrt{E_1 E_2} \lesssim \exp 10^{18}$ and it can be large.

\subsubsection{Rear-end particle collision with $\sigma_1 \sigma_2=1$}
For a rear-end particle collision, we set $\sigma_1 \sigma_2=1$.
In the rear-end particle case, 
the colliding particles have large velocities in a same direction, 
and the relative velocity is small. 
Thus, the center-of-mass energy of the collision of the rear-end particles becomes small. 
The center-of-mass energy of the rear-end collision at the throat $r=r_0=2M$ 
is obtained as 
\begin{eqnarray}
E_{cm}^2(r_0)
&\sim&m_1^2+m_2^2 -\frac{L_1 L_2}{2M^2} \nonumber\\
&&+\left( m_1^2 +\frac{L_1^2}{4M^2} \right)\frac{E_2}{E_1}
+\left( m_2^2 +\frac{L_2^2}{4M^2} \right)\frac{E_1}{E_2} \nonumber\\
\end{eqnarray}
and it is small unless $m_i$, $\left| L_i \right| /M$, $E_1/E_2$, or $E_2/E_1$ is large.

\subsection{Circular orbit particle case}
The Damour-Solodukhin wormhole spacetime has a circular orbit at the throat $\rho=0$ or $r=M$ as shown below.
The existence of the circular orbit is a remarkable feature in the Damour-Solodukhin wormhole spacetime 
because such a circular orbit does not exist at $r=2M$ in the Schwarzschild spacetime.
From 
\begin{eqnarray}
v(0)=\left. \frac{dv}{d\rho} \right|_{\rho=0}=0,
\end{eqnarray}
the circular orbit exists at the throat $\rho=0$ for a particle with a fine-tuned angular momentum $L=L_c$, 
where $L_c$ is given by
\begin{eqnarray}
L_c \equiv \pm 2M \sqrt{\frac{E^2}{\Lambda^2}-m^2}. 
\end{eqnarray}
The second derivative of the effective potential $v(\rho)$ with respect to $\rho$ for the particle with a circular motion at the throat $\rho=0$ with the conserved angular momentum $L=L_c$
is given by 
\begin{eqnarray}
v''(0)\equiv \left. \frac{d^2v}{d\rho^2} \right|_{\rho=0}=\frac{e^2-2e^2\Lambda^2+2\Lambda^4}{8m^2M^2\Lambda^4}.
\end{eqnarray}

In the case $e\geq \sqrt{2}$, 
the circular orbit is stable $(v''(0)>0)$ for $0<\Lambda < \Lambda_-$ or $\Lambda_+ < \Lambda$,
it is marginally stable $(v''(0)=0)$ for $\Lambda = \Lambda_{\pm}$,  
and it is unstable $(v''(0)< 0)$ for $\Lambda_- < \Lambda < \Lambda_+$.  
Here, $\Lambda_{\pm}$ are defined by
\begin{eqnarray}
\Lambda_{\pm}\equiv \frac{\sqrt{2e\left( e \pm \sqrt{e^2-2} \right)}}{2}.
\end{eqnarray}
In the other case, $e<\sqrt{2}$,
the circular orbit is stable for $0<\Lambda$.

We have an interest in the wormhole with small $\Lambda \ll 1$ and a marginal particle with $e \sim O(1)$.
In this case, the circular orbit at the throat is stable, 
and the absolute value of the conserved angular momentum of the particle 
\begin{eqnarray}
\left|L\right|=\left|L_c\right| \sim \frac{2ME}{\Lambda}
\end{eqnarray} 
must be large.

We consider that either of two particles rotates at the throat and the other particle falls into the throat and they collide there.
The particle with the circular orbit does not move in the radial direction, i.e., the $\rho$ direction, 
and the other particle has a large velocity in the radial direction. 
Therefore, the relative velocity of the particles is large, 
and the center-of-mass energy of the collision of the particles becomes large as given by 
\begin{eqnarray}\label{eq:E_cm_cir} 
\frac{E_{cm}(r_0)}{\sqrt{E_1 E_2}}
\sim \frac{2}{\Lambda}.
\end{eqnarray}

\section{General, static, and spherically symmetric wormhole spacetime}
In this section, we generalize the center-of-mass energy of two particles in the Damour-Solodukhin wormhole spacetime obtained in Sec.~II-B 
to the one in a general, static, and spherically symmetric wormhole spacetime. 
Then, for examples of the wormhole spacetime, we consider the center-of-mass energy in a Reissner-Nordstr\"{o}m black-hole-like wormhole~\cite{Lemos:2008cv}, 
Krasnikov's wormhole ~\cite{Krasnikov:2018nga}, and Ellis wormhole spacetime~\cite{Ellis_1973,Bronnikov_1973}.

The line element in the general, static, and spherically symmetric wormhole spacetime is expressed by
\begin{eqnarray}\label{eq:line_element12} 
ds^2
=-A(r)dt^2 +B(r)dr^2 +r^2 \left( d \theta^2 +\sin^2\theta d \phi^2 \right), \nonumber\\
\end{eqnarray}
where $A(r)$ is a positive function and a wormhole throat exists at $r=r_0$, which satisfies $1/B(r_0)=0$. 
From a straightforward calculation, we get the center-of-mass energy $E_{cm}$ of two particles in the form
\begin{eqnarray}\label{eq:Ccm_general0} 
E_{cm}^2(r)
&=&m_1^2+m_2^2+\frac{2 E_1 E_2 }{A(r)} \nonumber\\
&&-2  \sigma_1 \sigma_2 B(r) \sqrt{V_1(r)V_2(r)} -\frac{2 L_1 L_2}{r^2} \nonumber\\
&=&m_1^2+m_2^2+\frac{2 E_1 E_2 }{A(r)}-\frac{2 L_1 L_2}{r^2} \nonumber\\
&&-2  \sigma_1 \sigma_2 \sqrt{ m_1^2- \frac{E_1^2}{A(r)} +\frac{L_1^2}{r^2}} \nonumber\\
&&\times \sqrt{ m_2^2- \frac{E_2^2}{A(r)} +\frac{L_2^2}{r^2}},
\end{eqnarray}
where the effective potential $V_i(r)$ for the radial motion of particle $i=1$ or $2$ in a radial coordinate $r$ is given by
\begin{eqnarray}\label{eq:Vi2} 
V_i(r) = \frac{1}{B(r)} \left( m_i^2- \frac{E_i^2}{A(r)} +\frac{L_i^2}{r^2} \right). 
\end{eqnarray}
We notice that the center-of-mass energy $E_{cm}$ does not depend on $B(r)$.
Introducing a proper radial coordinate~$\rho$ given by
\begin{eqnarray}\label{eq:rho_general} 
\frac{d\rho}{dr}=\sqrt{B(r)},
\end{eqnarray}
the center-of-mass energy $E_{cm}$ is expressed as
\begin{eqnarray}\label{eq:Ccm_generalv} 
E_{cm}^2(r)
&=&m_1^2+m_2^2+\frac{2 E_1 E_2 }{A(r)} \nonumber\\
&&-2  \sigma_1 \sigma_2 \sqrt{v_1(r)v_2(r)} -\frac{2 L_1 L_2}{r^2},
\end{eqnarray}
where $v_i(r)\equiv v_{i}(\rho(r))$ is the effective potential of particle $i=1$ or $2$ given by
\begin{eqnarray}
v_{i}(r)=v_{i}(\rho(r))= m_i^2- \frac{E_i^2}{A(r)} +\frac{L_i^2}{r^2},
\end{eqnarray} 
for the motion in the $\rho$ direction. 

If $\sqrt{A(r_0)} \ll E_1/m_1$, $E_2/m_2$, $E_1r_0/\left| L_1 \right|$, and $E_2r_0/\left| L_2 \right|$ are satisfied,
the center-of-mass energy of a head-on collision with $\sigma_1 \sigma_2=-1$ at the throat $r=r_0$
is given by
\begin{eqnarray}\label{eq:Ccm_general} 
\frac{E_{cm}(r_0)}{\sqrt{E_1 E_2}}
\sim \frac{2}{\sqrt{A(r_0)}}.
\end{eqnarray}

\subsection{Reissner-Nordstr\"{o}m black-hole-like wormhole~\cite{Lemos:2008cv}}
We consider a Reissner-Nordstr\"{o}m black-hole-like wormhole with 
\begin{eqnarray}\label{eq:line_element3} 
A(r)&=&f(r)+\Lambda^2, \\
B(r)&=&\frac{1}{f(r)}, \\
f(r)&=&1-\frac{2M}{r}+\frac{Q^2}{r^2},
\end{eqnarray}
where $Q$ is a constant and we assume $-M \leq Q \leq M$.
The wormhole was investigated by Lemos and Zaslavskii~\cite{Lemos:2008cv}.
The wormhole becomes the Damour-Solodukhin wormhole when $Q=0$.
The throat is at $r=r_0=M+\sqrt{M^2-Q^2}$.
Under the assumption $\Lambda \ll E_1/m_1$, $E_2/m_2$, $E_1r_0/\left| L_1 \right|$, and $E_2r_0/\left| L_2 \right|$, 
the center-of-mass energy $E_{cm}$ of the head-on collision at the throat $r=r_0$ is given by
\begin{eqnarray}\label{eq:Ecm_our} 
\frac{E_{cm}(r_0)}{\sqrt{E_1 E_2}}
\sim \frac{2}{\Lambda}.
\end{eqnarray}
This is the same as the center-of-mass energy~(\ref{eq:Ecm_headon_DS}) in the Damour-Solodukhin wormhole spacetime with $Q=0$.
It can be arbitrarily large in a black hole limit~$\Lambda \rightarrow 0$ 
as well as the center-of-mass energy of a head-on particle collision in an extremal limit in a overcharged Reissner-Nordstr\"{o}m spacetime~\cite{Patil:2011uf}.
See Eq.~(18) in Ref~\cite{Patil:2011uf}.

In Ref.~\cite{Zaslavskii:2013nra}, Zaslavskii considered almost-extremal, stationary, and axisymmetric spacetime without an event horizon.
In a static case, the $(t,t)$ component of the metric tensor is given by 
\begin{eqnarray}
g_{tt}(r)=-C(r) \left[ (r-r_{\mathrm{col}})^2 + r_{\mathrm{col}}^2 \varepsilon^2 \right],
\end{eqnarray}
where $C(r)$ is a positive function, $\varepsilon$ is a small positive parameter $\varepsilon \ll 1$,
and $r_{\mathrm{col}}$ is a positive constant
\footnote{ Note that $C(r)$ and $r_{\mathrm{col}}$ are denoted by $B(r)$ and $r_0$, respectively, in Ref.~\cite{Zaslavskii:2013nra}, 
while $B(r)$ denotes the $(r,r)$ component of the metric tensor $g_{rr}(r)$, and $r_0$ denotes the position of a throat in this paper.}. 
The center-of-mass energy for the head-on collision of two particles at $r=r_{\mathrm{col}}$ is given by 
\begin{eqnarray}\label{eq:Ecm_Zaslavskii} 
\frac{E_{cm}(r_{\mathrm{col}})}{\sqrt{E_1 E_2}}\sim \frac{2}{\sqrt{C(r_{\mathrm{col}})}r_{\mathrm{col}} \varepsilon}.
\end{eqnarray}
See Eqs.~(1), (4), (11), and (13) in Ref.~\cite{Zaslavskii:2013nra}.

We can apply Zaslavskii's method for the near-extremal case with $Q=\pm M$ and with $\Lambda \ll 1$ only. 
In this case, from $C(r)$, $r_{\mathrm{col}}$, and $\varepsilon$, which are given by
\begin{eqnarray}
C(r)&=&\frac{1+\Lambda^2}{r^2},\\
r_{\mathrm{col}}&=&\frac{M}{1+\Lambda^2},
\end{eqnarray}
and
\begin{eqnarray}
\epsilon=\Lambda,
\end{eqnarray}
respectively, 
the center-of-mass energy~(\ref{eq:Ecm_Zaslavskii}) for the head-on collision at $r=r_{\mathrm{col}}$ is obtained as
\begin{eqnarray}
\frac{E_{cm}(r_{\mathrm{col}})}{\sqrt{E_1 E_2}}
\sim \frac{2}{\Lambda}.
\end{eqnarray}
We notice $r_{\mathrm{col}} \sim r_0=M$.
Thus, we have recovered $E_{cm}(r_0)/\sqrt{E_1 E_2}$ given by Eq.~(\ref{eq:Ecm_our}) by Zaslavskii's method.
See Zaslavskii~\cite{Zaslavskii:2013nra,Zaslavskii:2018kix} for the details of the head-on collision in the near-extremal case.

\subsection{Krasnikov's wormhole metric~\cite{Krasnikov:2018nga}}
Krasnikov considered a black-hole-like wormhole as a particle accelerator
with a high center-of-mass energy of a head-on collision of two particles~\cite{Krasnikov:2018nga}. 
The functions $A(r)$ and $B(r)$ are given by
\begin{eqnarray}\label{eq:line_element4} 
A(r)&=&e^{2h(r)}, \\
B(r)&=&\frac{1}{1-\frac{2M}{r}}, 
\end{eqnarray}
where
\begin{eqnarray}\label{eq:h(r)} 
h(r)&=&\frac{1}{2} \ln \left(1- \frac{2M}{r} \right) \qquad    \mathrm{for} \:\: 6M \leq r \\
h(r)&=&-\frac{1}{2} \left( k-\frac{1}{12} \right) \left( \frac{r}{2M} -3 \right)^2 +\frac{1}{12} \left( \frac{r}{2M} -3 \right) \nonumber\\
&&+\frac{1}{2} \ln \frac{2}{3}  \qquad  \mathrm{for} \:\: 4M \leq r < 6M \\
h(r)&=& \frac{kr}{2M} -\frac{5k}{2} -\frac{1}{24} +\frac{1}{2} \ln \frac{3}{2}  \qquad   \mathrm{for} \:\: 2M \leq r < 4M, \nonumber\\
\end{eqnarray}
where $k\gg 1$ is a positive constant and the wormhole throat is at $r=r_0=2M$.
Krasnikov considered the collision of head-on particles at the throat in a limit $k \rightarrow \infty$. 
We notice that the wormhole becomes a black-hole-like wormhole in the limit $k \rightarrow \infty$ because of $A(r_0)\rightarrow 0$. 
From Eq.~(\ref{eq:Ccm_general}) and $A(r_0)\rightarrow 0$, 
the center-of-mass energy of the head-on particle collision in the black hole limit $k \rightarrow \infty$ could diverge.

\subsection{Ellis wormhole~\cite{Ellis_1973,Bronnikov_1973}}
The Ellis wormhole filled with a phantom scalar field~\cite{Ellis_1973,Bronnikov_1973}
is the first and the simplest example of the Morris-Thorne wormhole~\cite{Morris_Thorne_1988}.
The phenomena of the Ellis wormhole, such as 
the deflection angle of light~\cite{Chetouani_Clement_1984,Nandi:2006ds,Dey_Sen_2008,Muller:2008zza,Gibbons_Vyska_2012,Bhattacharya:2010zzb,Nakajima_Asada_2012,Tsukamoto_Harada_Yajima_2012,Tsukamoto:2016jzh,Tsukamoto:2016qro,Nandi:2016uzg,Jusufi:2017gyu,Bhattacharya:2019kkb},
gravitational lensing~\cite{Perlick_2004_Phys_Rev_D,Muller:2008zza,Abe_2010,Toki_Kitamura_Asada_Abe_2011,Tsukamoto_Harada_Yajima_2012,Tsukamoto_Harada_2013,Yoo_Harada_Tsukamoto_2013,Izumi_2013,Nakajima:2014nba,Bozza:2015wbw,%
Lukmanova_2016,Tsukamoto:2016qro,Tsukamoto:2016zdu,Tsukamoto:2017edq,Bozza:2017dkv,Asada:2017vxl,Perlick:2014zwa,Tsukamoto:2017hva,Lukmanova:2018dwz,Shaikh:2019jfr},
visualizations~\cite{Muller_2004}, 
shadows~\cite{Ohgami:2015nra,Ohgami:2016iqm,Kuniyasu:2018cgv,Perlick:2015vta}, 
the quasinormal mode ~\cite{Konoplya:2016hmd,Konoplya:2018ala},
quantum metrology~\cite{Sabin:2016zqo,Sabin:2017dvx},
the uniqueness~\cite{Yazadjiev:2017twg},
a rotation curve~\cite{Bozza:2015haa}, and so on, 
have been investigated 
because of its simplicity.
The upper bound of the number density from observations~\cite{Takahashi_Asada_2013} has been also obtained.
Its instability~\cite{Shinkai_Hayward_2002} is known contrary to an earlier work~\cite{Armendariz-Picon_2002}.

Wormholes supported by various matter with the same metric as 
the Ellis wormhole have been investigated~\cite{Kar:2002xa,Das:2005un,Shatskiy:2008us,Novikov:2012uj,Myrzakulov:2015kda,Tamang:2015tmd,Mai:2017riq,Gibbons:2017jzk,Goulart:2017iko,Aygun:2018lga,Canate:2019spb,Canate:2019oly},
and the stability of the wormhole with electrically charged dust with negative energy density under linearly spherically symmetric and axial perturbations has been found~\cite{Bronnikov:2013coa}.

The Ellis wormhole metric has 
\begin{eqnarray}\label{eq:line_element8} 
A(r)&=&1, \\
B(r)&=&\frac{1}{1-\frac{a^2}{r^2}},
\end{eqnarray}
where $a$ is a positive constant and the throat is at $r=r_0=\pm a$.
From Eq.~(\ref{eq:Ccm_general0}), 
the center-of-mass energy is given by 
\begin{eqnarray}\label{eq:Ccm_Ellis} 
E_{cm}^2(r)
&=&m_1^2+m_2^2+ 2 E_1 E_2  -\frac{2 L_1 L_2}{r^2}\nonumber\\
&&-2  \sigma_1 \sigma_2 \sqrt{\left( m_1^2- E_1^2 +\frac{L_1^2}{r^2} \right)\left( m_2^2- E_2^2 +\frac{L_2^2}{r^2} \right)} \nonumber\\
\end{eqnarray}
and it is not large when $E_1$ and $E_2$ are not large as discussed in Ref.~\cite{Tsukamoto:2014swa}. 
If we use a proper radial distance $\rho$ given by $\left| \rho \right| = \sqrt{r^2-a^2}$,
the line element is rewritten as
\begin{eqnarray}\label{eq:line_element9} 
ds^2 =-dt^2 +d\rho^2 +\left( \rho^2+a^2 \right) \left( d \theta^2 +\sin^2\theta d \phi^2 \right),\nonumber\\
\end{eqnarray}
and the center-of-mass energy~(\ref{eq:Ccm_Ellis}) is rewritten as (A3) and (A4) in Ref.~\cite{Tsukamoto:2014swa}.

\section{Conclusion}
Damour and Solodukhin considered a wormhole with a metric which is similar to 
the Schwarzschild spacetime~\cite{Damour:2007ap}.
We have shown that the center-of-mass energy for the head-on collision of two particles in the Damour-Solodukhin wormhole spacetime can be large.
The center-of-mass energy is equal to an upper bound of the total mass of products after the particle collision.

We comment on differences between the collision of particles with the large center-of-mass energy in the Damour-Solodukhin wormhole spacetime and
the BSW collision in an extremal black hole spacetime~\cite{Banados:2009pr}.
In the Damour-Solodukhin wormhole, 
the head-on particles with the large center-of-mass energy do not have a critical angular momentum and
the particles reach into the throat in a finite proper time.
On the other hand, the BSW collision is a rear-end collision between a particle with a critical conserved angular momentum, 
and a particle with a noncritical conserved angular momentum near the extremal black hole. 
The particle with the critical angular momentum reaches into an extremal horizon in an infinite proper time.

We have found that the center-of-mass energy of the head-on particle collision at the throat of a general, static, and spherically symmetric wormhole 
can be large if the $(t,t)$ component of the metric tensor is similar to black holes.
One can apply our method for near-extremal and non-near-extremal cases, 
while Zaslavskii's method~\cite{Zaslavskii:2013nra} works in the near-extremal case only.
The behavior of the arbitrary high center-of-mass energy of the head-on particle collision at the throat of non-near-extremal wormholes in a black hole limit 
is the same as the near-extremal wormhole case at least in our approach.

The center-of-mass energy will be suppressed for the collision of celestial objects with finite sizes such as stars and planets. 
On the other hand, it can be large for the collision of fundamental particles like neutrons and photons,
and then a small black hole may be formed at the collisional point after the collision.
However, the small black hole will evaporate before it absorbs matters supporting the wormhole throat.
The finite-size effect of falling objects on the center-of-mass energy and the effect of the collision on the background spacetime
will depend the gravitational sector of the theory, the matters supporting the throat, and the falling objects.
We will need an analysis beyond a test-particle approximation to treat them.  
We hope that this paper stimulates researchers to study the details of the particle collisions near the wormhole.

\section*{Acknowledgements}
The authors are grateful to J.~P.~S. Lemos for bringing their attention to related work~\cite{Lemos:2008cv} when they were finalizing the present paper.
They thank an anonymous referee for valuable comments and suggestions.
They are grateful to R.~N.~Izmailov, K.~K.~Nandi, and O.~B.~Zaslavskii for valuable comments.
T.K. acknowledges the support for this work by the National Natural Science Foundation of China under Grant No.~11690034.
\appendix
\section{Ricci scalar}
In this Appendix, we comment on a relation of the center-of-mass energy of two particles and the Ricci scalar $R(r)$ in wormhole spacetimes.
The Ricci scalar $R(r)$ of the Damour-Solodukhin wormhole is obtained as 
\begin{eqnarray}
R(r)=-\frac{2 \Lambda^2 M^2}{r^2 \left(\Lambda^2 r-2 M+r\right)^2} 
\end{eqnarray}
and it is, on the throat $r=2M$, 
\begin{eqnarray}
R(2M)=-\frac{1}{8 \Lambda^{2} M^{2}}.
\end{eqnarray}
Thus, the absolute value of the Ricci scalar on the throat is large when $\Lambda$ is small. 

However, we have to keep in mind that a large Ricci scalar in wormhole spacetimes does not cause the high center-of-mass energy of the particles always as shown below.
The Ricci scalar $R(r)$ of the Ellis wormhole~\cite{Ellis_1973,Bronnikov_1973} is given by
\begin{eqnarray}
R(r)=-\frac{2 a^2}{r^4} 
\end{eqnarray}
and the Ricci scalar on the throat $r=a$ is obtained as 
\begin{eqnarray}
R(a)=-\frac{2}{a^2}.
\end{eqnarray}
Thus, the absolute value of the Ricci scalar $R(a)$ on the throat can be large if $a$ is small.
On the other hand, the center-of-mass energy~(\ref{eq:Ccm_Ellis}) of the collisions of two particles, 
which are initially far away from the throat and which do not have large conserved energy $E$,
cannot be large even if $a$ is small and $\left|R(a)\right|$ is large as discussed in Ref.~\cite{Tsukamoto:2014swa}.

\end{document}